\documentclass[11pt]{article}%
\usepackage{amsmath}
\usepackage{amsfonts}
\usepackage{amssymb}
\usepackage{fullpage}
\usepackage{graphicx}%
\begin{document}

\title{NP-complete Problems and Physical Reality}
\author{Scott Aaronson\thanks{Institute for Advanced Study, Princeton, NJ. \ Email:
aaronson@ias.edu. \ Supported by the NSF.}}
\date{}
\maketitle

\begin{abstract}
Can $\mathsf{NP}$-complete\ problems be solved efficiently in the physical
universe? \ I survey proposals including soap bubbles, protein folding,
quantum computing, quantum advice, quantum adiabatic algorithms,
quantum-mechanical nonlinearities, hidden variables, relativistic time
dilation, analog computing, Malament-Hogarth spacetimes, quantum gravity,
closed timelike curves, and ``anthropic computing.'' \ The section on soap
bubbles even includes some ``experimental'' results. \ While I do not believe
that any of the proposals will let us solve $\mathsf{NP}$-complete problems
efficiently, I argue that by studying them, we can learn something not only
about computation but also about physics.

\end{abstract}

\section{Introduction\label{INTRO}}

\begin{quotation}
\textquotedblleft Let a computer smear---with the right kind of quantum
randomness---and you create, in effect, a `parallel' machine with an
astronomical number of processors \ldots\ All you have to do is be sure that
when you collapse the system, you choose the version that happened to find the
needle in the mathematical haystack.\textquotedblright

---From \textit{Quarantine} \cite{egan}, a 1992 science-fiction novel by Greg Egan
\end{quotation}

If I had to debate the science writer John Horgan's claim that basic science
is coming to an end \cite{horgan}, my argument would lean heavily on one fact:
\textit{it has been only a decade since we learned that quantum computers
could factor integers in polynomial time}. \ In my (unbiased) opinion, the
showdown that quantum computing has forced---between our deepest intuitions
about computers on the one hand, and our best-confirmed theory of the physical
world on the other---constitutes one of the most exciting scientific dramas of
our time.

But why did this drama not occur until so recently? \ Arguably, the main ideas
were already in place by the 1960's or even earlier. \ I do not know the
answer to this sociological puzzle, but can suggest two
possibilities.\ \ First, many computer scientists see the study of
\textquotedblleft speculative\textquotedblright\ models of computation as at
best a diversion from more serious work; this might explain why the
groundbreaking papers of Simon \cite{simon}\ and Bennett et al.\ \cite{bbbv}%
\ were initially rejected from the major theory conferences. \ And second,
many physicists see computational complexity as about as relevant to the
mysteries of Nature as\ dentistry or tax law.

Today, however, it seems clear that there is something to gain from resisting
these attitudes. \ We would do well to ask: \textit{what} \textit{else} about
physics might we have overlooked in thinking about the limits of efficient
computation? \ The goal of this article is to encourage the serious discussion
of this question. \ For concreteness, I will focus on a single sub-question:
\textit{can }$\mathsf{NP}$\textit{-complete problems be solved in polynomial
time using the resources of the physical universe?}

I will argue that studying this question can yield new insights, not just
about computer science but about physics as well. \ More controversially, I
will also argue that a negative answer might eventually attain the same status
as (say) the Second Law of Thermodynamics, or the impossibility of
superluminal signalling. \ In other words, while experiment will always be the
last appeal, the presumed intractability of $\mathsf{NP}$-complete\ problems
might be taken as a useful constraint in the search for new physical theories.
\ Of course, the basic concept will be old hat to computer scientists who live
and die by the phrase, \textquotedblleft Assuming $\mathsf{P}\neq\mathsf{NP}$,
\ldots\textquotedblright

To support my arguments, I will survey a wide range of unusual computing
proposals, from soap bubbles and folding proteins to time travel, black holes,
and quantum nonlinearities. \ Some of the proposals are better known than
others, but to my knowledge, even the \textquotedblleft
folklore\textquotedblright\ ones have never before been collected in one
place. \ In evaluating the proposals, I will try to insist that \textit{all}
relevant resources be quantified, and \textit{all} known physics taken into
account. \ As we will see, these straightforward ground rules have been
casually ignored in some of the literature on\ exotic computational models.

Throughout the article, I assume basic familiarity with complexity classes
such as $\mathsf{P}$ and $\mathsf{NP}$ (although not much more than that).
\ Sometimes I do invoke elementary physics concepts, but the difficulty of the
physics is limited by my own ignorance.

After reviewing the basics of $\mathsf{P}$\ versus $\mathsf{NP}$\ in Section
\ref{BASICS}, I discuss soap bubbles and related proposals in Section
\ref{SOAP}, and even report some original \textquotedblleft
experimental\textquotedblright\ work in this field. \ Then Section
\ref{QUANTUM} summarizes what is known about solving $\mathsf{NP}$-complete
problems on a garden-variety quantum computer; it includes discussions of
black-box lower bounds, quantum advice, and the quantum adiabatic algorithm.
Section \ref{VQM}\ then considers \textit{variations} on quantum mechanics
that might lead to a more powerful model of computation; these include
nonlinearities in the Schr\"{o}dinger equation\ and certain assumptions about
hidden variables. \ Section \ref{REL} moves on to consider analog computing,
time dilation, and exotic spacetime geometries;\ this section is basically a
plea to those who think about these matters, to take seriously such
trivialities as quantum mechanics and the Planck scale. \ Relativity and
quantum mechanics finally meet in Section\ \ref{QG}, on the computational
complexity of quantum gravity theories, but the whole point of the section is
to explain why this is a premature subject. \ Sections \ref{TIME} and
\ref{ANTHROPIC} finally set aside the more sober ideas (like solving the
halting problem using naked singularities), and give zaniness free reign.
\ Section \ref{TIME}\ studies the computational complexity of time travel,
while Section \ref{ANTHROPIC} studies \textquotedblleft anthropic
computing,\textquotedblright\ which means killing yourself whenever a computer
fails to produce a certain output. \ It turns out that even about these
topics, there are nontrivial things to be said! \ Finally, Section\ \ref{DISC}
makes the case for taking the hardness of $\mathsf{NP}$-complete\ problems to
be a basic fact about the physical world; and weighs three possible objections
against doing so.

I regret that, because of both space and cognitive limitations, I was unable
to discuss \textit{every} paper related to the solvability of $\mathsf{NP}%
$-complete problems\ in the physical world. \ Two examples of
omissions are the gear-based computers of Vergis, Steiglitz, and
Dickinson \cite{vsd}, and the proposed adiabatic algorithm for the
halting problem due to Kieu \cite{kieu}. \ Also, I generally ignored
papers about \textquotedblleft hypercomputation\textquotedblright\
that did not try to forge \textit{some} link, however tenuous, with
the laws of physics as we currently understand them.

\section{The Basics\label{BASICS}}

I will not say much about the original $\mathsf{P}$\ versus $\mathsf{NP}$
question: only that the known heuristic algorithms for the 3SAT problem, such
as backtrack, simulated annealing, GSAT, and survey propagation, can solve
some instances quickly in practice, but are easily stumped by other instances;
that the standard opinion is that $\mathsf{P}\neq\mathsf{NP}$ \cite{gasarch};
that proving this is correctly seen as one of the deepest problems in all of
mathematics \cite{clay}; that no one has any idea where to begin
\cite{feinstein}; and that we have a pretty sophisticated idea of \textit{why}
we have no idea \cite{rr}. \ See \cite{sipser:pnp} or \cite{gj}\ for more information.

Of course, even if there is no deterministic algorithm to solve $\mathsf{NP}%
$-complete problems in polynomial time, there might be a probabilistic
algorithm, or a nonuniform algorithm (one that is different for each input
length $n$). \ But Karp and Lipton \cite{kl} showed that either of these would
have a consequence, namely the collapse of the polynomial hierarchy, that
seems almost as implausible as $\mathsf{P}=\mathsf{NP}$. \ Also, Impagliazzo
and Wigderson \cite{iw}\ gave strong evidence that $\mathsf{P}=\mathsf{BPP}$;
that is, that any probabilistic algorithm can be simulated by a deterministic
one with polynomial slowdown.

It is known that $\mathsf{P}\neq\mathsf{NP}$\ in a \textquotedblleft black
box\textquotedblright\ or oracle setting \cite{bgs}. \ This just means that
any efficient algorithm for an $\mathsf{NP}$-complete problem would have to
exploit the problem's structure in a nontrivial way, as opposed to just trying
one candidate solution after another until it finds one that works.
\ Interestingly, most of the physical proposals for solving $\mathsf{NP}%
$-complete problems\ that we will see do \textit{not} exploit structure, in
the sense that they would still work relative to any oracle. \ Given this
observation, I propose the following challenge: \textit{find a physical
assumption under which }$\mathsf{NP}$\textit{-complete\ problems can provably
be solved in polynomial time, but only in a non-black-box setting.}

\section{Soap Bubbles et al.\label{SOAP}}

\begin{figure}
[ptb]
\begin{center}
\includegraphics[
trim=2.311020in 3.079452in 1.927918in 3.083593in,
height=1.7036in,
width=1.4346in
]%
{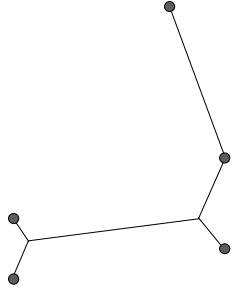}%
\caption{A Steiner tree connecting points at $(.7,.96)$, $(.88,.46)$,
$(.88,.16)$, $(.19,.26)$, $(.19,.06)$ (where $\left(  0,0\right)  $\ is in the
lower left corner, and $\left(  1,1\right)  $\ in the upper right). \ There
are two Steiner vertices, at roughly $\left(  .24,.19\right)  $ and\ $\left(
.80,.26\right)  $.}%
\label{steiner}%
\end{center}
\end{figure}

Given a set of points in the Euclidean plane, a \textit{Steiner
tree} (see Figure \ref{steiner}) is a collection of line segments of
minimum total length connecting the points, where the segments can
meet at vertices (called Steiner vertices) other than the points
themselves. \ Garey, Graham, and Johnson
\cite{ggj} showed that finding such a tree is $\mathsf{NP}$%
-hard.\footnote{Naturally, the points' coordinates must be specified to some
finite precision. \ If we only need to decide whether there exists a tree of
total length at most $L$, or whether all trees have length at least
$L+\varepsilon$\ (for some small $\varepsilon>0$), then the problem becomes
$\mathsf{NP}$-complete.} \ Yet a well-known piece of computer science folklore
maintains that, if two glass plates with pegs between them are dipped into
soapy water, then the soap bubbles will rapidly form a Steiner tree connecting
the pegs, this being the minimum-energy configuration.%
%TCIMACRO{\FRAME{ftbpFU}{1.4346in}{1.7036in}{0pt}{\Qcb{A Steiner tree
%connecting points at $(.7,.96)$, $(.88,.46)$, $(.88,.16)$, $(.19,.26)$,
%$(.19,.06)$ (where $\left(  0,0\right)  $\ is in the lower left corner, and
%$\left(  1,1\right)  $\ in the upper right). \ There are two Steiner vertices,
%at roughly $\left(  .24,.19\right)  $ and\ $\left(  .80,.26\right)  $.}%
%}{\Qlb{steiner}}{steiner.eps}{\special{ language "Scientific Word";
%type "GRAPHIC";  maintain-aspect-ratio TRUE;  display "USEDEF";
%valid_file "F";  width 1.4346in;  height 1.7036in;  depth 0pt;
%original-width 7.7551in;  original-height 10.3511in;  cropleft "0.2980";
%croptop "0.7021";  cropright "0.7514";  cropbottom "0.2975";
%filename 'steiner.eps';file-properties "XNPEU";}}}%
%BeginExpansion
%EndExpansion

It was only a matter of time before \textit{someone} put the pieces together.
\ Last summer Bringsjord and Taylor \cite{bt} posted a paper entitled
\textquotedblleft P=NP\textquotedblright\ to the arXiv. \ This paper argues
that, since (1) finding a Steiner tree is $\mathsf{NP}$-hard, (2) soap bubbles
find a Steiner tree in polynomial time, (3) soap bubbles are classical
objects, and (4) classical physics can be simulated by a Turing machine with
polynomial slowdown, it follows that $\mathsf{P}=\mathsf{NP}$.

My immediate reaction was that the paper was a parody. \ However, a visit to
Bringsjord's home page\footnote{www.rpi.edu/\symbol{126}brings} \ suggested
that it was not. \ Impelled, perhaps, by the same sort of curiosity that
causes people to watch reality TV shows, I checked the discussion of this
paper on the comp.theory newsgroup\ to see if anyone recognized the obvious
error. \ And indeed, several posters pointed out that, although soap bubbles
might reach a minimum-energy configuration with a small number of pegs, there
is no \textquotedblleft magical\textquotedblright\ reason why this should be
true in general. \ By analogy, a rock in a mountain crevice could reach a
lower-energy configuration by rolling up first and then down, but it is not
observed to do so. \ A poster named Craig Feinstein replied to these skeptics
as follows \cite{feinstein:soap}:

\begin{quote}
Have experiments been done to show that it is only a local minimum that is
reached by soap bubbles and not a global minimum or is this just the party
line? \ I'd like to believe that nature was designed to be smarter than we
give it credit. \ I'd be willing to make a gentleman's bet that no one can
site [sic] a paper which describes an experiment that shows that the global
minimum is not always achieved with soap bubbles.
\end{quote}

Though I was unable to find such a paper, I was motivated by this post to
conduct the experiment myself.\footnote{S. Aaronson,\ NP-complete Problems and
Physical Reality, \textit{SIGACT\ News} Complexity Theory Column, March 2005.
\ I win.} \ I bought two $8\textquotedblright\times9\textquotedblright$ glass
plates, paint to mark grid points on the plates, thin copper rods which I cut
into $1\textquotedblright$ pieces, suction cups to attach the rods to the
plates, liquid oil soap, a plastic tub to hold the soapy water, and work
gloves. \ I obtained instances of the Euclidean Steiner Tree problem from the
OR-Library website \cite{beasley}. \ I concentrated on instances with $3$ to
$7$ vertices, for example the one shown in Figure \ref{steiner}.

The result was fascinating to watch: with $3$ or $4$ pegs, the optimum tree
usually \textit{is} found. \ However, by no means is it always found,
especially with more pegs. \ Soap-bubble partisans might write this off as
experimental error, caused (for example) by inaccuracy in placing the pegs, or
by the interference of my hands. \ However, I also sometimes found triangular
\textquotedblleft bubbles\textquotedblright\ of three Steiner vertices---which
is much harder to explain, since such a structure could never occur in a
Steiner tree. \ In general, the results were highly nondeterministic; I could
obtain entirely different trees by dunking the same configuration more than
once. \ Sometimes I even obtained a tree that did not connect all the pegs.

Another unexpected phenomenon was that sometimes the bubbles would start in a
suboptimal configuration, then slowly \textquotedblleft
relax\textquotedblright\ toward a better one. \ Even with $4$ or $5$ pegs,
this process could take around ten seconds, and it is natural to predict that
with more pegs it would take longer. \ In short, then, I found no reason to
doubt the \textquotedblleft party line,\textquotedblright\ that soap bubbles
do not solve $\mathsf{NP}$-complete\ problems in polynomial time by
magic.\footnote{Some people have objected that, while all of this might be
true in practice, I still have not shown that soap bubbles cannot solve
$\mathsf{NP}$-complete\ problems \textit{in principle}. \ But what exactly
does \textquotedblleft in principle\textquotedblright\ mean? \ If it means
obeying the equations of classical physics, then the case for magical
avoidance of local optima moves from empirically weak to demonstrably false,
as in the case of the rock stuck in the mountain crevice.}

There are other proposed methods for solving $\mathsf{NP}$-complete\
problems that involve relaxation to a minimum-energy state, such as
spin glasses \ and protein folding. \ All of these methods are
subject to the same pitfalls of local optima and potentially long
relaxation times. \ Protein folding is an interesting case, since it
seems likely that proteins evolved specifically \textit{not} to have
local optima. \ A protein that folded in unpredictable ways could
place whatever organism relied on it at an adaptive disadvantage
(although sometimes it happens anyway, as with prions). \ However,
this also means that if we engineered an artificial protein to
represent a hard 3SAT\ instance, then there would be no particular
reason for it to fold as quickly or reliably as do naturally
occurring proteins.

\section{Quantum Computing\label{QUANTUM}}

Outside of theoretical computer science, parallel computers are sometimes
discussed as if they were fundamentally more powerful than serial computers.
\ But of course, anything that can be done with $10^{20}$\ processors in time
$T$ can also be done with one processor in time $10^{20}T$. \ The same is true
for DNA strands. \ Admittedly, for some applications a constant factor of
$10^{20}$\ is not irrelevant.\footnote{This is one fact I seem to remember
from my computer architecture course.} \ But for solving (say) 3SAT instances
with hundreds of thousands of variables, $10^{20}$\ is peanuts.

When quantum computing came along, it was hoped that finally we might have a
type of parallelism commensurate with the difficulty of $\mathsf{NP}%
$-complete\ problems. \ For in quantum mechanics, we need a vector of $2^{n}%
$\ complex numbers\ called \textquotedblleft amplitudes\textquotedblright%
\ just to specify the state of an $n$-bit computer (see
\cite{aar:thesis,fortnow:qc,nc} for more details). \ Surely we could exploit
this exponentiality inherent in Nature to try out all $2^{n}$\ possible
solutions to an $\mathsf{NP}$-complete\ problem in parallel? \ Indeed, many
popular articles on quantum computing have given precisely that impression.

The trouble is that if we measure the computer's state, we see only
\textit{one} candidate solution $x$, with probability depending on
its amplitude $\alpha_{x}$.\footnote{Some authors recognized this
difficulty even in the 1980's; see Pitowsky \cite{pitowsky} for
example.} \ The challenge is to arrange the computation in such a
way that only the $x$'s we wish to see wind up with large values of
$\alpha_{x}$. \ For the special case of factoring, Shor \cite{shor}
showed that this could be done using a polynomial number of
operations---but what about for $\mathsf{NP}$-complete\ problems?

The short answer is that we don't know. \ Indeed, letting $\mathsf{BQP}$\ be
the class of problems solvable in polynomial time by a quantum computer, we do
not even know whether $\mathsf{NP}\subseteq\mathsf{BQP}$ would imply
$\mathsf{P}=\mathsf{NP}$\ or some other unlikely consequence in classical
complexity.\footnote{On the other hand, if $\mathsf{\#P}$-complete problems
were solvable in quantum polynomial time, then this \textit{would} have an
unlikely classical complexity consequence, namely the collapse of the
so-called counting hierarchy.} \ But in 1994, Bennett, Bernstein, Brassard,
and Vazirani \cite{bbbv} did show that\ $\mathsf{NP}\not \subset \mathsf{BQP}%
$\ relative to an oracle.\ \ In particular, they showed that any quantum
algorithm that searches an unordered database of $N$\ items for a single
\textquotedblleft marked\textquotedblright\ item must query the database
$\sim\sqrt{N}$ times.\ \ (Soon afterward, Grover \cite{grover}\ showed that
this is tight.)

If we interpret the space of $2^{n}$\ possible assignments to a Boolean
formula $\varphi$\ as a \textquotedblleft database,\textquotedblright\ and the
satisfying assignments of $\varphi$\ as \textquotedblleft marked
items,\textquotedblright\ then Bennett et al.'s result says that any quantum
algorithm needs at least $\sim2^{n/2}$\ steps to find a satisfying assignment
of $\varphi$\ with high probability, \textit{unless} the algorithm exploits
the structure of $\varphi$\ in a nontrivial way. \ In other words, there is no
\textquotedblleft brute-force\textquotedblright\ quantum algorithm to solve
$\mathsf{NP}$-complete\ problems\ in polynomial time, just as there is no
brute-force classical algorithm.

In Bennett et al.'s original proof, we first run our quantum algorithm on a
database with no marked items. \ We then mark the item that was queried with
the smallest total probability, and show that the algorithm will need many
queries to notice this change. \ By now, many other\ proofs have been
discovered, including that of Beals et al.\ \cite{bbcmw}, which represents an
efficient quantum algorithm's acceptance probability by a low-degree
polynomial, and then shows that no such polynomial exists; and that of
Ambainis \cite{ambainis}, which upper-bounds how much the entanglement between
the algorithm and database can increase via a single query. \ Both techniques
have also led to lower bounds for many other problems besides database search.

The crucial property of quantum mechanics that all three proofs exploit is its
\textit{linearity}: the fact that, until a measurement is made, the vector of
amplitudes can only evolve by means of linear transformations. \ Intuitively,
if we think of the components of a superposition as \textquotedblleft parallel
universes,\textquotedblright\ then linearity is what prevents the universe
containing the marked item from simply \textquotedblleft telling all the other
universes about it.\textquotedblright

\subsection{Quantum Advice\label{ADVICE}}

The above assumed that our quantum computer begins in some standard initial
state, such as the \textquotedblleft all-$0$\textquotedblright\ state (denoted
$\left\vert 0\cdots0\right\rangle $). \ An interesting twist is to consider
the effects of other initial states. \ Are there quantum states that could
take exponential time to prepare, but that would let us solve $\mathsf{NP}%
$-complete\ problems\ in polynomial time\ were they given to us by a wizard?
\ More formally, let $\mathsf{BQP/qpoly}$\ be the class of problems solvable
in quantum polynomial time, given a polynomial-size \textquotedblleft quantum
advice state\textquotedblright\ $\left\vert \psi_{n}\right\rangle $\ that
depends only on the input length $n$.\ Then recently I showed that
$\mathsf{NP}\not \subset \mathsf{BQP/qpoly}$ relative to an oracle
\cite{aar:adv}. \ Intuitively, even if the state $\left\vert \psi
_{n}\right\rangle $ encoded the solutions to every 3SAT instance of size $n$,
only a miniscule fraction of that information could be extracted by measuring
$\left\vert \psi_{n}\right\rangle $, at least within the black-box model that
we know how to analyze. \ The proof uses the polynomial technique of Beals et
al.\ \cite{bbcmw}\ to prove a so-called \textit{direct product theorem}, which
upper-bounds the probability of solving many database search problems
simultaneously. \ It then shows that this direct product theorem could be
violated, if the search problem were efficiently solvable using quantum advice.

\subsection{The Quantum Adiabatic Algorithm\label{ADIABAT}}

At this point, some readers may be getting impatient with the black-box model.
\ After all, $\mathsf{NP}$-complete\ problems\ are \textit{not} black boxes,
and classical algorithms such as backtrack search do exploit their structure.
\ Why couldn't a quantum algorithm do the same? \ A few years ago, Farhi et
al.\ \cite{fggllp} announced a new \textit{quantum adiabatic algorithm}, which
can be seen as a quantum analogue of simulated annealing. \ Their algorithm is
easiest to describe in a continuous-time setting, using the concepts of a
Hamiltonian (an operation that acts on a quantum state over an infinitesimal
time interval $\Delta t$) and a ground state (the lowest-energy state left
invariant by a given Hamiltonian). \ The algorithm starts by applying a
Hamiltonian $H_{0}$\ that has a known, easily prepared ground state, then
slowly transitions to another Hamiltonian $H_{1}$\ whose ground state encodes
the solution to (say) an instance of 3SAT. \ The \textit{quantum adiabatic
theorem} says that if a quantum computer starts in the ground state of $H_{0}%
$, then it must end in the ground state of $H_{1}$, \textit{provided the
transition from }$H_{0}$\textit{\ to }$H_{1}$\textit{\ is slow enough}. \ The
key question is how slow is slow enough.

In their original paper, Farhi et al.\ \cite{fggllp} gave numerical evidence
that the adiabatic algorithm solves random, critically-constrained instances
of the $\mathsf{NP}$-complete\ Exact Cover problem in polynomial time. \ But
having learned from experience, most computer scientists are wary of taking
such numerical evidence too seriously as a guide to asymptotic behavior.
\ This is especially true when the instance sizes are small ($n\leq20$ in
Farhi et al.'s case), as they have to be when simulating a quantum computer on
a classical one. \ On the other hand, Farhi relishes pointing out that if the
empirically-measured running time were exponential, no computer scientist
would dream of saying that it would eventually become polynomial! \ In my
opinion, the crucial experiment (which has not yet been done) would be to
compare the adiabatic algorithm head-on against simulated annealing and other
classical heuristics. \ The evidence for the adiabatic algorithm's
performance\ would be much more convincing if the known classical algorithms
took exponential time on the same random instances.

On the theoretical side, van Dam, Mosca, and Vazirani \cite{dmv} constructed
3SAT instances for which the adiabatic algorithm provably takes exponential
time, at least when the transition between the initial and final Hamiltonians
is linear. \ Their instances involve a huge \textquotedblleft basin of
attraction\textquotedblright\ that leads to a false optimum (meaning most but
not all clauses are satisfied), together with an exponentially small basin
that leads to the true optimum. \ To lower-bound the algorithm's running time
on these instances, van Dam et al.\ showed that the spectral gap (that is, the
gap between the smallest and second-smallest eigenvalues) of some intermediate
Hamiltonian decreases exponentially in $n$. \ As it happens, physicists have
almost a century of experience in analyzing these spectral gaps, but not for
the purpose of deciding whether they decrease polynomially or exponentially as
the number of particles increases to infinity!

Such \textquotedblleft hands-on\textquotedblright\ analysis of the adiabatic
algorithm was necessary, since van Dam et al.\ also showed that there is no
black-box proof that the algorithm takes exponential time. \ This is because,
given a variable assignment $X$ to the 3SAT\ instance $\varphi$, the adiabatic
algorithm computes not merely \textit{whether} $X$ satisfies $\varphi$, but
also how many clauses it satisfies. \ And this information turns out to be
sufficient to reconstruct $\varphi$ itself.

Recently Reichardt \cite{reichardt}, building on work of Farhi, Goldstone, and
Gutmann \cite{fgg}, has constructed 3SAT instances for which the adiabatic
algorithm takes polynomial time, whereas simulated annealing takes exponential
time. \ These instances involve a narrow obstacle along the path to the global
optimum, which simulated annealing gets stuck at but which the adiabatic
algorithm tunnels past. \ On the other hand, these instances are easily solved
by other classical algorithms. \ An interesting open question is whether there
exists a family of black-box functions $f:\left\{  0,1\right\}  ^{n}%
\rightarrow\mathbb{Z}$\ for which the adiabatic algorithm finds a global
minimum using exponentially fewer queries than \textit{any} classical algorithm.

\section{Variations on Quantum Mechanics\label{VQM}}

Quantum computing skeptics sometimes argue that we do not \textit{really} know
whether quantum mechanics itself will remain valid in the regime tested by
quantum computing.\footnote{Personally, I agree, and consider this the main
motivation for trying to build a quantum computer.} \ Here, for example, is
Leonid Levin \cite{levin:qc}: \textquotedblleft The major problem [with
quantum computing] is the requirement that basic quantum equations hold to
multi-hundredth if not millionth decimal positions where the significant
digits of the relevant quantum amplitudes reside. \ We have never seen a
physical law valid to over a dozen decimals.\textquotedblright

The irony is that most of the specific proposals for how quantum mechanics
\textit{could} be wrong suggest a world with more, not less, computational
power than $\mathsf{BQP}$. \ For, as we saw in Section \ref{QUANTUM}, the
linearity of quantum mechanics is what prevents one needle in an exponentially
large haystack from shouting above the others. \ And as observed by Weinberg
\cite{weinberg}, it seems difficult to change quantum mechanics in any
consistent way while preserving linearity.

But how drastic could the consequences possibly be, if we added a tiny
nonlinear term to the Schr\"{o}dinger equation (which describes how quantum
states evolve in time)? \ For starters, Gisin \cite{gisin}\ and Polchinski
\cite{polchinski}\ showed that in most nonlinear variants of quantum
mechanics, one could use entangled states to transmit superluminal signals.
\ More relevant for us, Abrams and Lloyd \cite{al}\ showed that one could
solve $\mathsf{NP}$-complete and even $\mathsf{\#P}$-complete\ problems in
polynomial time---at least if the computation were error-free. \ Let us see
why this is, starting with $\mathsf{NP}$.

Given a black-box function $f$ that maps $\left\{  0,1\right\}  ^{n}$\ to
$\left\{  0,1\right\}  $, we want to decide in polynomial time whether there
exists an input $x$ such that $f\left(  x\right)  =1$. \ We can start by
preparing a uniform superposition over all inputs, denoted $2^{-n/2}\sum
_{x}\left\vert x\right\rangle $, and then querying the oracle for $f$, to
produce $2^{-n/2}\sum_{x}\left\vert x\right\rangle \left\vert f\left(
x\right)  \right\rangle $. \ If we then apply Hadamard gates to the
first\ register and measure that register, one can show that we will obtain
the outcome $\left\vert 0\cdots0\right\rangle $\ with probability at least
$1/4$. \ Furthermore, conditioned on the first register having the state
$\left\vert 0\cdots0\right\rangle $, the second register will be in the state%
\[
\frac{\left(  2^{n}-s\right)  \left\vert 0\right\rangle +s\left\vert
1\right\rangle }{\sqrt{\left(  2^{n}-s\right)  ^{2}+s^{2}}}%
\]
where $s$ is the number of inputs $x$ such that $f\left(  x\right)  =1$. \ So
the problem reduces to that of distinguishing two possible states of a single
qubit---for example, the states corresponding to $s=0$\ and $s=1$. \ The only
difficulty is that these states are exponentially close.

But a nonlinear operation need not preserve the angle between quantum
states---it can pry them apart. \ Indeed, Abrams and Lloyd showed that by
repeatedly applying a particular kind of nonlinear gate, which arises in a
model of Weinberg \cite{weinberg}, one could increase the angle between two
quantum states \textit{exponentially}, and thereby distinguish the $s=0$\ and
$s=1$\ cases with constant bias. \ It seems likely that \textquotedblleft
almost any\textquotedblright\ nonlinear gate would confer the same ability,
though it is unclear how to formalize this statement.

To solve $\mathsf{\#P}$-complete problems, we use the same basic algorithm,
but apply it repeatedly to \textquotedblleft zoom in\textquotedblright\ on the
value of $s$ using binary search. \ Given any range $\left[  a,b\right]  $
that we believe contains $s$, by applying the nonlinear gate a suitable number
of times we can make the case $s=a$\ correspond roughly to $\left\vert
0\right\rangle $, and the case $s=b$\ correspond roughly to $\left\vert
1\right\rangle $. \ Then measuring the state will provide information about
whether $s$ is closer to $a$ or $b$. \ This is true even if $\left(
b-a\right)  /2^{n}$\ is exponentially small.

Indeed, if arbitrary $1$-qubit\ nonlinear operations are allowed,\ then it is
not hard to see that we could even solve $\mathsf{PSPACE}$-complete\ problems
in polynomial time. \ It suffices to solve the following problem: given a
Boolean function $f$ of $n$ bits $x_{1},\ldots,x_{n}$, does there exist a
setting of $x_{1}$\ such that for all settings of $x_{2}$ there exists a
setting of $x_{3}$\ such that\ldots\ $f\left(  x_{1},\ldots,x_{n}\right)  =1$?
\ To solve this, we can first prepare the state%
\[
\frac{1}{2^{n/2}}\sum_{x_{1},\ldots,x_{n}}\left\vert x_{1}\ldots
x_{n},f\left(  x_{1}\ldots x_{n}\right)  \right\rangle .
\]
We then apply a \textquotedblleft nonlinear AND\ gate\textquotedblright\ to
the $n^{th}$\ and $\left(  n+1\right)  ^{st}$\ qubits,\ which maps $\left\vert
00\right\rangle +\left\vert 10\right\rangle $, $\left\vert 00\right\rangle
+\left\vert 11\right\rangle $, and $\left\vert 01\right\rangle +\left\vert
10\right\rangle $ to $\left\vert 00\right\rangle +\left\vert 10\right\rangle
$, and $\left\vert 01\right\rangle +\left\vert 11\right\rangle $\ to itself
(omitting the $\sqrt{2}$ normalization). \ Next we apply a \textquotedblleft
nonlinear OR gate\textquotedblright\ to the $\left(  n-1\right)  ^{st}$\ and
$\left(  n+1\right)  ^{st}$\ qubits,\ which maps $\left\vert 00\right\rangle
+\left\vert 11\right\rangle $, $\left\vert 01\right\rangle +\left\vert
10\right\rangle $, and $\left\vert 01\right\rangle +\left\vert 11\right\rangle
$ to $\left\vert 01\right\rangle +\left\vert 11\right\rangle $, and
$\left\vert 00\right\rangle +\left\vert 10\right\rangle $\ to itself. \ We
continue to alternate between AND\ and OR in this manner, while moving the
control qubit leftward towards $x_{1}$. \ At the end, the $\left(  n+1\right)
^{st}$\ qubit will be $\left\vert 1\right\rangle $\ if the answer is `yes,'
and $\left\vert 0\right\rangle $\ if the answer is `no.'

On the other hand, any nonlinear quantum computer can also be simulated in
$\mathsf{PSPACE}$. \ For even in nonlinear theories, the amplitude of any
basis state at time $t$ is an easily-computable function of a small number of
amplitudes at time $t-1$, and can therefore be computed in polynomial space
using depth-first recursion. \ It follows that, assuming arbitrary nonlinear
gates and no error, $\mathsf{PSPACE}$\ exactly characterizes the power of
nonlinear quantum mechanics.

But what if we allow error, as any physically reasonable model of computation
must? \ In this case, while it \textit{might} still be possible to
solve\ $\mathsf{NP}$-complete problems in polynomial time, I am not convinced
that Abrams and Lloyd have demonstrated this.\footnote{Abrams and Lloyd
claimed to give an algorithm that does not require \textquotedblleft
exponentially precise operations.\textquotedblright\ \ The problem is that
their algorithm uses a nonlinear OR gate, and depending on how that gate
behaves on states other than $\left\vert 00\right\rangle +\left\vert
10\right\rangle $, $\left\vert 00\right\rangle +\left\vert 11\right\rangle $,
$\left\vert 01\right\rangle +\left\vert 10\right\rangle $, and $\left\vert
01\right\rangle +\left\vert 11\right\rangle $, it might magnify small errors
exponentially. \ In particular, I could not see how to implement a nonlinear
OR\ gate robustly using Abrams and Lloyd's \textquotedblleft Weinberg
gate.\textquotedblright} \ Observe that the standard quantum error-correction
theorems break down, since just as a tiny probability of success can be
magnified exponentially during the course of a computation, so too can a tiny
probability of error. \ Whether this problem can be overcome might depend on
which specific nonlinear gates are available; the issue deserves further investigation.

\subsection{Hidden-Variable Theories\label{HIDDEN}}

Most people who quote Einstein's declaration that \textquotedblleft God does
not play dice\textquotedblright\ seem not to realize that a dice-playing God
would be an \textit{improvement} over the actual situation. \ In quantum
mechanics, a particle does not have a position, even an unknown position,
until it is measured. \ This means that it makes no sense to talk about a
\textquotedblleft trajectory\textquotedblright\ of the particle, or even a
probability distribution over possible trajectories. \ And without such a
distribution, it is not clear how we can make even probabilistic predictions
for future observations, if we ourselves belong to just one component of a
larger superposition.

Hidden-variable theories try to remedy this problem by supplementing quantum
mechanics with the \textquotedblleft actual\textquotedblright\ values of
certain observables (such as particle positions or momenta), together with
rules for how those observables evolve in time. \ The most famous such theory
is due to Bohm \cite{bohm}, but there are many alternatives that are equally
compatible with experiment. \ Indeed, a key feature of hidden-variable
theories is that they reproduce the usual quantum-mechanical probabilities\ at
any individual time, and so are empirically indistinguishable from ordinary
quantum mechanics. \ It does not seem, therefore, that a \textquotedblleft
hidden-variable quantum computer\textquotedblright\ could possibly be more
powerful than a garden-variety one.

On the other hand, it might be that Nature needs to \textquotedblleft expend
more computational effort\textquotedblright\ to calculate a particle's entire
trajectory than to calculate its position at any individual time. \ The reason
is that the former requires keeping track of multiple-time correlations. \ And
indeed, I showed in \cite{aar:qchv}\ that under any hidden-variable theory
satisfying a reasonable axiom called \textquotedblleft indifference to the
identity,\textquotedblright\ the ability to sample the hidden variable's
history would let us solve the Graph Isomorphism problem in polynomial time.
\ For intuitively, given two graphs $G$ and $H$ with no nontrivial
automorphisms, one can easily prepare a uniform superposition over all
permutations of $G$ and $H$:%
\[
\frac{1}{\sqrt{2n!}}\sum_{\sigma\in S_{n}}\left(  \left\vert 0\right\rangle
\left\vert \sigma\right\rangle \left\vert \sigma\left(  G\right)
\right\rangle +\left\vert 1\right\rangle \left\vert \sigma\right\rangle
\left\vert \sigma\left(  H\right)  \right\rangle \right)  .
\]
Then measuring the third register yields a state of the form $\left\vert
i\right\rangle \left\vert \sigma\right\rangle $\ if $G$ and $H$ are not
isomorphic, or $\left(  \left\vert 0\right\rangle \left\vert \sigma
\right\rangle +\left\vert 1\right\rangle \left\vert \tau\right\rangle \right)
/\sqrt{2}$\ for some $\sigma\neq\tau$\ if they are isomorphic.
\ Unfortunately, if then we measured this state in the standard basis, we
would get no information whatsoever, and work of myself \cite{aar:col}, Shi
\cite{shi},\ and Midrijanis \cite{midrijanis}\ shows that no
black-box\ quantum algorithm can do much better. \ But if only we could make a
few \textquotedblleft non-collapsing\textquotedblright\ measurements! \ Then
we would see the same permutation each time in the former case, but two
permutations with high probability in the latter.

The key point is that seeing a hidden variable's history would effectively let
us simulate non-collapsing measurements. \ Using this fact, I showed\ that by
sampling histories, we could simulate the entire class $\mathsf{SZK}$\ of
problems having statistical zero-knowledge proofs, which includes Graph
Isomorphism, Approximate Shortest Vector, and other $\mathsf{NP}$-intermediate
problems for which no efficient quantum algorithm is known. \ On the other
hand, $\mathsf{SZK}$\ is not thought to contain the $\mathsf{NP}$-complete
problems; indeed, if it did then the polynomial hierarchy would collapse
\cite{bhz}. \ And it turns out that, even if we posit the unphysical ability
to sample histories, we \textit{still} could not solve $\mathsf{NP}$-complete
problems efficiently in the black-box setting! \ The best we could do is
search a list of $N$ items in $\sim N^{1/3}$\ steps, as opposed to $\sim
N^{1/2}$\ with Grover's algorithm.

But even if a hidden-variable picture is correct, are these considerations
relevant to any computations \textit{we} could perform? \ They would be, if a
proposal of Valentini \cite{valentini,valentini:thesis} were to pan out.
\ Valentini argues that the $\left\vert \psi\right\vert ^{2}$\ probability law
merely reflects a statistical equilibrium (analogous to thermal equilibrium),
and that it might be possible to find \textquotedblleft nonequilibrium
matter\textquotedblright\ (presumably left over from the Big Bang) in which
the hidden variables still obey a different distribution. \ Using such matter,
Valentini showed that we could distinguish nonorthogonal states, and thereby
transmit superluminal signals and break quantum cryptographic protocols. \ He
also claimed that we could solve $\mathsf{NP}$-complete\ problems in
polynomial time. \ Unfortunately, his algorithm involves measuring a
particle's position to exponential precision, and if we could do that, then it
is unclear why we could not also solve $\mathsf{NP}$-complete\ problems in
polynomial time \textit{classically}! \ So in my view, the power of
Valentini's model with realistic constraints on precision remains an
intriguing open question. \ My conjecture is that it will turn out to be
similar to the power of the histories model---that is, able to solve
$\mathsf{SZK}$\ problems in polynomial time,\ but not $\mathsf{NP}$-complete
problems\ in the black-box setting. \ I would love to be disproven.

\section{Relativity and Analog Computing\label{REL}}

If quantum computers cannot solve $\mathsf{NP}$-complete problems efficiently,
then perhaps we should turn to the other great theory of twentieth-century
physics: relativity. \ The idea of relativity computing is simple: start your
computer working on an intractable problem, then board a spaceship and
accelerate to nearly the speed of light. \ When you return to Earth, all of
your friends will be long dead, but the answer to your problem will await you.

What is the problem with this proposal? \ Ignoring the time spent accelerating
and decelerating, if you travelled at speed $v$ relative to Earth for proper
time $t$ (where $v=1$ is light speed), then the elapsed time in your
computer's reference frame would be $t^{\prime}=t/\sqrt{1-v^{2}}$. \ It
follows that, if you want $t^{\prime}$\ to increase exponentially with $t$,
then $v$ has to be exponentially close to the speed of light. \ But this
implies that the amount of \textit{energy} needed to accelerate the spaceship
also increases exponentially with $t$. \ So your spaceship's fuel tank (or
whatever else is powering it) will need to be exponentially large---which
means that you will again need exponential time, just for the fuel from the
far parts of the tank to affect you!

Similar remarks apply to traveling close to a black hole event
horizon: if you got exponentially close then you would need
exponential energy to escape.\footnote{An interesting property of
relativity is that it is always \textit{you} who has to go somewhere
or do something in these proposals, while the computer stays behind.
\ Conversely, if you wanted more time to think about what to say
next in a conversation, then your conversational partner is the one
who would have to be placed in a spaceship.} \ On the other hand,
Malament and Hogarth (see \cite{hogarth}) have constructed
spacetimes in which, by traveling for a finite proper time along one
worldline, an observer could see the \textit{entire infinite past}
of another worldline. Naturally, this would allow that observer to
solve not only $\mathsf{NP}$-complete problems\ but the halting
problem as well. \ It is known that these spacetimes cannot be
globally hyperbolic; for example, they could have naked
singularities, which are points at which general relativity no
longer yields predictions. \ But to me, the mere existence of such
singularities is a relatively minor problem, since there is evidence
today that they really \textit{can} form in classical general
relativity (see \cite{singh} for a survey).

The real problem is the Planck scale. \ By combining three physical
constants---Planck's constant $\hbar\approx1.05\times10^{-34}m^{2}kg^{1}%
s^{-1}$, Newton's gravitational constant $G\approx6.67\times10^{-11}%
m^{3}kg^{-1}s^{-2}$, and the speed of light $c\approx3.00\times10^{8}%
m^{1}kg^{0}s^{-1}$---one can obtain a fundamental unit of length known as the
\textit{Planck length}:%
\[
\ell_{P}=\sqrt{\frac{\hbar G}{c^{3}}}\approx1.62\times10^{-35}m.
\]
The physical interpretation of this length is that, if we tried to confine an
object inside a sphere of diameter $\ell_{P}$, then the object would acquire
so much energy that it would collapse to form a black hole. \ For this reason,
most physicists consider it meaningless to discuss lengths shorter than the
Planck length, or times shorter than the corresponding Planck time $\ell
_{P}/c\approx5.39\times10^{-44}s$. \ They assume that, even if there do exist
naked singularities, these are simply places where\ general relativity breaks
down on length scales of order $\ell_{P}$, and must be replaced by a quantum
theory of gravity.

Indeed, Bekenstein \cite{bekenstein}\ gave an upper bound on the total
information content of any isolated, weakly gravitating physical system,\ by
assuming the Second Law of Thermodynamics and then considering a thought
experiment in which the system is slowly lowered into a black hole.
\ Specifically, he showed that $S\leq2\pi ER$, where $S$ is the entropy of the
system, or $\ln2$ times the number of bits of information; $E$ is the system's
gravitating energy; and $R$ is the radius of the smallest sphere containing
the system. \ Note that $E$ and $R$ are in Planck units. \ Since the energy of
a system can be at most proportional to its radius (at least according to the
widely-believed \textquotedblleft hoop conjecture\textquotedblright), one
corollary of Bekenstein's bound is the \textit{holographic bound}: the
information content of any region is at most proportional to the surface area
of the region, at a rate of one bit per Planck length squared, or
$1.4\times10^{69}$\ bits per square meter. \ Bousso \cite{bousso}, whose
survey paper on this subject is well worth reading by computer
scientists,\ has reformulated the holographic bound in a generally covariant
way, and marshaled a surprising amount of evidence for its validity.

Some physicists go even further, and maintain that space and time are
literally discrete on the Planck scale. \ Of course, the discreteness could
not be of the straightforward kind that occurs in cellular automata such as
Conway's Game of Life, since that would fail to reproduce Lorentz or even
Galilean invariance. \ Instead, it would be a more subtle, quantum-mechanical
kind of discreteness, as appears for example\ in loop quantum gravity (see
Section \ref{QG}). \ But I should stress that the holographic bound itself,
and the existence of a Planck scale at which classical ideas about space and
time break down, are generic conclusions that stand independently of any
specific quantum gravity theory.

The reason I have taken this detour into Planck-scale physics\ is that our
current understanding seems to rule out, not only the Malament-Hogarth
proposal, but \textit{all similar proposals} for solving the halting problem
in finite time. Yet in the literature on \textquotedblleft
hypercomputation\textquotedblright\ \cite{copeland,hogarth}, one still reads
about machines that could \textquotedblleft bypass the Turing
barrier\textquotedblright\ by performing the first step of a computation in
one second, the second in $1/2$\ second, the third in $1/4$\ second, and so
on, so that after two seconds an infinite number of steps has been performed.
\ Sometimes the proposed mechanism invokes Newtonian physics (ignoring even
the finiteness of the speed of light), while other times it requires traveling
arbitrarily close to a spacetime singularity. \ Surprisingly, in the papers
that I encountered, the most common response to quantum effects was not to
discuss them at all!

The closest I found to an account of physicality comes from Hogarth
\cite{hogarth}, who stages an interesting dialogue between a traditional
computability theorist named Frank and a hypercomputing enthusiast named
Isabel. \ After Isabel describes a type of spacetime that would support
\textquotedblleft non-Turing computers,\textquotedblright\ the following
argument ensues:

\begin{quote}
\textit{Frank:} Yes, but surely the spacetime underlying our universe is not
like that. \ These solutions [to Einstein's equation] are just idealisations.

\textit{Isabel:} That's beside the point. \ You don't want to rubbish a
hypothetical computer---Turing or non-Turing---simply because it can't fit
into our universe. \ If you do, you'll leave your precious Turing machine to
the mercy of the cosmologists, because according to one of their theories, the
universe and all it contains, will crunch to nothing in a few billion years.
\ Your Turing machine would be cut-off in mid-calculation! \cite[p.
15]{hogarth}
\end{quote}

I believe that Isabel's analogy fails. \ For in principle, one can generally
translate theorems about Turing machines into statements about what Turing
computers could or could not do\ within the time and space bounds of the
physical universe.\footnote{As an example, Stockmeyer and Meyer \cite{sm}%
\ gave a simple problem in logic, such that solving instances of size $610$
provably requires circuits with at least $10^{125}$ gates.} \ By contrast, it
is unclear if claims about hypercomputers have \textit{any} relevance
whatsoever to the physical universe. \ The reason is that, if the $n^{th}$
step of a hypercomputation took $2^{-n}$\ seconds, then it would take fewer
than $150$ steps to reach the Planck time.

In my view, the \textquotedblleft foaminess\textquotedblright\ of space and
time on the Planck scale also rules out approaches to$\ \mathsf{NP}$-complete
problems based on analog computing. \ (For present purposes, an analog
computer is a machine that performs a discrete sequence of steps, but on
unlimited-precision real numbers.) \ As an example of such an approach, in
1979 Sch\"{o}nhage \cite{schonhage}\ showed how to solve $\mathsf{NP}%
$-complete\ and even $\mathsf{PSPACE}$-complete\ problems in polynomial time,
given the ability to compute $x+y$, $x-y$, $xy$,\ $x/y$, and $\left\lfloor
x\right\rfloor $ in a single time step for any two real numbers $x$\ and
$y\neq0$. \ Intuitively, one can use the first $2^{\Theta\left(  n\right)  }$
bits in a real number's binary expansion\ to encode an instance of the
Quantified Boolean Formula problem, then use arithmetic operations to
calculate the answer in parallel, and finally extract the binary
result.\footnote{Note that the ability to apply the floor function (or
equivalently, to access a specific bit in a real number's binary expansion) is
essential here. \ If we drop that ability, then we obtain the beautiful theory
of algebraic complexity \cite{bcss,bcs},\ which has its own \textquotedblleft%
$\mathsf{P}$ versus $\mathsf{NP}$\textquotedblright\ questions over the real
and complex numbers. \ These questions are logically unrelated to the original
$\mathsf{P}$ versus $\mathsf{NP}$ question so far as anyone knows---possibly
they are easier.} \ The problem, of course, is that unlimited-precision real
numbers would violate the holographic entropy bound.

\section{Quantum Gravity\label{QG}}

Here we enter a realm of dragons, where speculation abounds but concrete ideas
about computation are elusive. \ The one clear result is due to Freedman,
Kitaev, Larsen, and Wang \cite{fkw,flw}, who studied topological quantum field
theories (TQFT's). \ These theories, which arose from the work of Witten and
others in the 1980's,\ involve $2$ spatial dimensions and $1$ time dimension.
\ Dropping from $3$ to $2$ dimensions might seem like a trivial change to a
computer scientist, but it has the effect of making quantum gravity radically
simpler; basically, the only degree of freedom is now the topology of the
spacetime manifold, together with any \textquotedblleft
punctures\textquotedblright\ in that manifold. \ Surprisingly, Freedman et
al.\ were able to define a model of computation based on TQFT's, and show that
this model is equivalent to ordinary quantum computation: more precisely, all
TQFT's can be simulated in $\mathsf{BQP}$, and some TQFT's are universal for
$\mathsf{BQP}$. \ Unfortunately, the original papers on this discovery are all
but impossible for a computer scientist to read,\ but Aharonov, Jones, and
Landau \cite{ajl} are currently working on a simplified presentation.

From what I understand, it remains open to analyze the computational
complexity of $\left(  3+1\right)  $-dimensional quantum field
theories even in flat spacetime. \ Part of the problem is that these
theories are not mathematically rigorous: they have well-known
infinities, which are swept under the rug via a process called
\textquotedblleft renormalization.\textquotedblright\ \ However,
since the theories in some sense preserve quantum-mechanical
unitarity, the expectation of physicists I have asked is that they
will not lead to a model of computation more powerful than
$\mathsf{BQP}$.

The situation is different for speculative theories incorporating gravity,
such as M-theory, the latest version of string theory. \ For these theories
involve a notion of \textquotedblleft locality\textquotedblright\ that is much
more subtle than the usual one:\ in particular, the so-called AdS/CFT
correspondence proposes that theories with gravity in $d$ dimensions are
somehow isomorphic to theories without gravity in $d-1$\ dimensions (see
\cite{deboer}). \ As a result, Preskill \cite{preskill:talk}\ has pointed out
that even if M-theory remains based on standard quantum mechanics, it might
allow the efficient implementation of unitary transformations that would
require exponential time on an ordinary quantum computer. \ It would be
interesting to develop this idea further.

String theory's main competitor is a theory called loop quantum
gravity.\footnote{If some physicist wants to continue the tradition of naming
quantum gravity theories using monosyllabic words for elongated objects that
mean something completely different in computer science, then I propose the
most revolutionary advance yet: \textit{thread theory}.} \ Compared to string
theory, loop quantum gravity has one feature that I find attractive as a
computer scientist: it explicitly models spacetime as discrete and
combinatorial on the Planck scale. \ In particular, one can represent the
states in this theory by sums over \textit{spin networks}, which are
undirected graphs with edges labeled by integers. \ The spin networks evolve
via local operations called Pachner moves; a sequence of these moves is called
a spin foam. \ Then the \textquotedblleft amplitude\textquotedblright\ for
transitioning from spin network $A$ to spin network $B$ equals the sum, over
all spin foams $F$ going from $A$ to $B$, of the amplitude of $F$. \ In a
specific model known as the Riemannian\footnote{Here \textquotedblleft
Riemannian\textquotedblright\ means not taking into account that time is
different from space. \ There is also a Lorentzian Barrett-Crane model, but it
is considerably more involved.} Barrett-Crane model, this amplitude equals the
product, over all Pachner moves in $F$, of an expression called a
\textquotedblleft10j symbol,\textquotedblright\ which can be evaluated
according to rules originally developed by Penrose \cite{penrose:spin}.

Complicated, perhaps, but this seems like the stuff out of which a
computational model could be made. \ So two years ago I spoke with Dan
Christensen, a mathematician who along with Greg Egan gave an efficient
algorithm \cite{ce}\ for calculating 10j symbols that has been crucial in the
numerical study of spin foams. \ I wanted to know whether one could define a
complexity class \textquotedblleft$\mathsf{BQGP}$\textquotedblright%
\ (Bounded-Error Quantum Gravity Polynomial-Time) based on spin foams, and if
so, how it compared to $\mathsf{BQP}$. \ The first observation we made is that
evaluating arbitrary spin networks (as opposed to 10j symbols) using Penrose's
rules is $\mathsf{\#P}$-complete. \ This follows by a simple reduction from
counting the number of edge 3-colorings of a trivalent planar graph, which was
proven $\mathsf{\#P}$-complete\ by Vertigan and Welsh \cite{vw}.

But what about simulating the dynamics of (say) the Barrett-Crane model?
\ Here we quickly ran into problems: for example, in summing over all spin
foams between two spin networks, should one impose an upper bound on the
number of Pachner moves, and if so, what? \ Also, supposing we \textit{could}
compute amplitudes for transitioning from one spin network to another, what
would these numbers represent? \ If they are supposed to be analogous to
transition amplitudes in ordinary quantum mechanics, then how do we normalize
them so that probabilities sum to unity? \ In the quantum gravity literature,
issues such as these are still not settled.\footnote{If the normalization were
done manually, then presumably\ one could solve $\mathsf{NP}$-complete
problems in polynomial time using postselection (see Section \ref{ANTHROPIC}).
\ This seems implausible.}

In the early days of quantum mechanics, there was much confusion about the
operational meaning of the wavefunction. \ (Even in Born's celebrated 1926
paper \cite{born}, the idea that one has to \textit{square} amplitudes to get
probabilities only appeared in a footnote added in press!) \ Similarly,
Einstein struggled for years to extract testable physics from a theory in
which any coordinate system is as valid as any other. \ So maybe it is no
surprise that, while today's quantum gravity researchers can write down
equations, they are still debating what seem to an outsider like extremely
basic questions about what the equations mean. \ The trouble is that these
questions are exactly the ones we need answered, if we want to formulate a
model of computation! \ Indeed, to anyone who wants a test or benchmark for a
favorite quantum gravity theory,\footnote{That is, one without all the bother
of making numerical predictions and comparing them to observation.} let me
humbly propose the following: \textit{can you define Quantum Gravity
Polynomial-Time?}

A possible first step would be to define time. \ For in many quantum gravity
theories, there is not even a notion of objects evolving dynamically in time:
instead there is just a static spacetime manifold, subject to a constraint
such as the Wheeler-DeWitt equation $H\Psi= 0$. \ In classical general
relativity, at least we could carve the universe into `spacelike slices' if we
wanted to, and assign a local time to any given observer! \ But how do we do
either of those if the spacetime metric itself is in quantum superposition?
\ Regulars call this ``the problem of time'' (see \cite{smolin:time} for a
fascinating discussion). \ The point I wish to make is that, until this and
the other conceptual problems have been clarified---until we can say what it
means for a `user' to specify an `input' and `later' receive an
`output'---\textit{there is no such thing as computation, not even
theoretically}.

\section{Time Travel Computing\label{TIME}}

Having just asserted that a concept of time something like the usual one is
needed even to define computation, I am now going to disregard that principle,
and discuss computational models that exploit closed timelike curves (CTC's).
\ The idea was well explained by the movie \textit{Star Trek IV: The Voyage
Home}. \ The Enterprise crew has traveled back in time to the present (meaning
to 1986) in order to find humpback whales and bring them into the twenty-third
century. \ The problem is that building a tank to transport the whales
requires a type of plexiglass that has not yet been invented. \ In
desperation, the crew seeks out the company that \textit{will} invent the
plexiglass, and reveals its molecular formula to that company. \ The question
is, where did the work of inventing the formula take place?

\begin{figure}
[ptb]
\begin{center}
\includegraphics[
trim=2.696462in 3.084979in 2.314506in 0.000000in,
height=1.6621in,
width=1.8979in
]%
{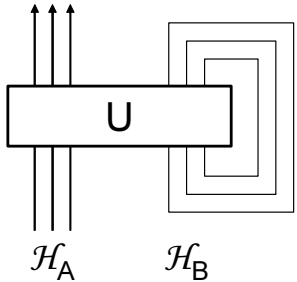}%
\caption{Deutsch's causal consistency model consists of `chronology-respecting
qubits' in Hilbert space $\mathcal{H}_{A}$, and `CTC qubits' in Hilbert space
$\mathcal{H}_{B}$\ whose quantum state must be invariant under $U$.}%
\label{ttfig}%
\end{center}
\end{figure}

In a classic paper on CTC's, Deutsch \cite{deutsch:tt}\ observes that, in
contrast to the much better-known grandfather paradox, the \textquotedblleft
knowledge creation paradox\textquotedblright\ involves no logical
contradiction. \ The only paradox\ is a complexity-theoretic one: a difficult
computation somehow gets performed, yet without the expected resources being
devoted to it. \ Deutsch goes further, and argues that this\ is \textit{the}
paradox of time travel, the other ones vanishing once quantum mechanics is
taken into account. \ The idea is this: consider a unitary matrix $U$ acting
on the Hilbert space $\mathcal{H}_{A}\otimes\mathcal{H}_{B}$, where
$\mathcal{H}_{A}$\ consists of `chronology-respecting\ qubits' and
$\mathcal{H}_{B}$\ consists of `closed timelike curve\ qubits' (see Figure
\ref{ttfig}). \ Then one can show that there always exists a mixed quantum
state $\rho$\ of the $\mathcal{H}_{B}$ qubits, such that if we start with
$\left\vert 0\cdots0\right\rangle $\ in $\mathcal{H}_{A}$ and $\rho$\ in the
$\mathcal{H}_{B}$, apply $U$, and then trace out $\mathcal{H}_{A}$, the
resulting state in $\mathcal{H}_{B}$\ is again $\rho$. \ Deutsch calls this
requirement \textit{causal consistency}. \ What it means is that $\rho$\ is a
\textit{fixed point} of the superoperator\footnote{A \textquotedblleft
superoperator\textquotedblright\ is a generalization of a unitary matrix that
can include interaction with ancilla qubits, and therefore need not be
reversible.} acting on $\mathcal{H}_{B}$, so we can take it to be both the
`input' and `output' of the CTC.%
%TCIMACRO{\FRAME{ftbpFU}{1.8979in}{1.6621in}{0pt}{\Qcb{Deutsch's causal
%consistency model consists of `chronology-respecting qubits' in Hilbert space
%$\mathcal{H}_{A}$, and `CTC qubits' in Hilbert space $\mathcal{H}_{B}$\ whose
%quantum state must be invariant under $U$.}}{\Qlb{ttfig}}{ttfig.eps}%
%width 1.8979in;  height 1.6621in;  depth 0pt;  original-width 10.3511in;
%original-height 7.7551in;  cropleft "0.2605";  croptop "1";
%cropright "0.7764";  cropbottom "0.3978";
%filename 'ttfig.eps';file-properties "XNPEU";}}}%
%BeginExpansion
%EndExpansion

Strictly speaking, Deutsch's idea does not depend on quantum mechanics; we
could equally well say that any Markov chain has a stationary distribution.
\ In both the classical and quantum cases, the resolution of the grandfather
paradox is then that you are born with $1/2$ probability, and \textit{if} you
are born you go back in time to kill your grandfather, from which it follows
that you are born with $1/2$ probability, and so on.

One advantage of this resolution is that it immediately suggests a model of
computation. \ For simplicity, let us first consider the classical case, and
assume \textit{all} bits go around the CTC (this assumption will turn out not
to matter for complexity purposes). \ Then the model is the following: first
the user specifies as input a polynomial-size circuit $C:\left\{  0,1\right\}
^{n}\rightarrow\left\{  0,1\right\}  ^{n}$. \ Then Nature chooses a
probability distribution $\mathcal{D}$\ over $\left\{  0,1\right\}  ^{n}%
$\ that is left invariant by $C$. \ Finally, the user receives as output a
sample $x$ from $\mathcal{D}$, which can be used as the basis for further
computation. \ An obvious question is, if there is more than one stationary
distribution $\mathcal{D}$, then which one does Nature choose? \ The answer
turns out to be irrelevant, since we can construct circuits $C$ such that a
sample from \textit{any} stationary distribution could be used to solve
$\mathsf{NP}$-complete or even $\mathsf{PSPACE}$-complete\ problems in
polynomial time.

The circuit for $\mathsf{NP}$-complete\ problems is simple: given a Boolean
formula $\varphi$, let $C\left(  x\right)  =x$\ if $x$ is a satisfying
assignment for $\varphi$, and $C\left(  x\right)  =x+1$\ otherwise, where $x$
is considered as an $n$-bit integer and the addition is mod $2^{n}$. \ Then
provided $\varphi$\ has any satisfying assignments at all, the only stationary
distributions of $C$\ will be the singleton distributions concentrated on
those assignments.

I am indebted to Lance Fortnow for coming up with a time travel circuit for
the more general case of $\mathsf{PSPACE}$-complete\ problems. \ Let
$M_{1},\ldots,M_{T}$\ be the successive configurations of a $\mathsf{PSPACE}%
$\ machine $M$. \ Then our circuit $C$ will take as input a machine
configuration $M_{t}$\ together with a bit $i\in\left\{  0,1\right\}  $. \ The
circuit does the following: if $t<T$, then $C$ maps each $\left(
M_{t},i\right)  $ to $\left(  M_{t+1},i\right)  $. \ Otherwise, if $t=T$, then
$C$\ maps $\left(  M_{T},i\right)  $\ to $\left(  M_{1},0\right)  $\ if
$M_{T}$\ is a rejecting state, or $\left(  M_{T},i\right)  $\ to $\left(
M_{1},1\right)  $\ if $M_{T}$\ is an accepting state. \ Notice that if $M$
accepts, then the only stationary distribution of $C$ is the uniform
distribution over the cycle $\left\{  \left(  M_{1},1\right)  ,\ldots,\left(
M_{T},1\right)  \right\}  $. \ On the other hand, if $M$ rejects, then the
only stationary distribution is uniform over $\left\{  \left(  M_{1},0\right)
,\ldots,\left(  M_{T},0\right)  \right\}  $. \ So in either case, measuring
$i$ yields the desired output.

Conversely, it is easy to see that a $\mathsf{PSPACE}$\ machine can sample
from some stationary distribution of $C$. \ For the problem reduces to finding
a cycle in the exponentially large graph of the function $C:\left\{
0,1\right\}  ^{n}\rightarrow\left\{  0,1\right\}  ^{n}$, and then choosing a
uniform random vertex from that cycle. \ The same idea works even if not all
$n$ of the bits go around the CTC. \ It follows that $\mathsf{PSPACE}%
$\ exactly characterizes the classical computational complexity of time
travel, if we assume Deutsch's causal consistency requirement.

But what about the \textit{quantum} complexity of time travel? \ The model is
as follows: first the user specifies a polynomial-size quantum circuit
$C$\ acting on $\mathcal{H}_{A}\otimes\mathcal{H}_{B}$; then Nature
adversarially chooses a mixed state $\rho$\ such that $\operatorname*{Tr}%
_{A}\left[  C\left(  \left\vert 0\cdots0\right\rangle \left\langle
0\cdots0\right\vert \otimes\rho\right)  \right]  =\rho$, where
$\operatorname*{Tr}_{A}$\ denotes partial trace over $\mathcal{H}_{A}$; and
finally the user can perform an arbitrary $\mathsf{BQP}$ computation on $\rho
$. \ Let $\mathsf{BQP}_{\mathsf{CTC}}$\ be the class of problems solvable in
this model. \ Then it is easy to see that $\mathsf{BQP}_{\mathsf{CTC}}$
contains $\mathsf{PSPACE}$, since we can simulate the classical time travel
circuit for $\mathsf{PSPACE}$\ using a quantum circuit. \ On the other hand,
the best \textit{upper} bound I know of on $\mathsf{BQP}_{\mathsf{CTC}}$\ is a
class called $\mathsf{SQG}$\ (Short Quantum Games), which was defined by
Gutoski and Watrous \cite{gw} and which generalizes $\mathsf{QIP}$\ (the class
of problems having quantum interactive proof protocols). \ Note that
$\mathsf{QIP}$\ contains but is not known to equal $\mathsf{PSPACE}$.
\ Proving that $\mathsf{BQP}_{\mathsf{CTC}}\subseteq\mathsf{SQG}$, and
hopefully improving on that result to pin down the power of $\mathsf{BQP}%
_{\mathsf{CTC}}$ exactly, are left as exercises for the reader.

\subsection{The Algorithms of Bacon and Brun\label{BACONBRUN}}

My goal above was to explore the computational power of time travel in a
clear, precise, complexity-theoretic way. \ However, there are several other
perspectives on time travel computing; two were developed by Bacon
\cite{bacon}\ and Brun \cite{brun}.

I assumed before that we have access to only one CTC, but can send a
polynomial number of bits (or qubits) around that CTC. \ Bacon
considers a different model, in which we might be able to send only
\textit{one} bit around a CTC, but can use a polynomial number of
CTC's. \ It is difficult to say which model is the more reasonable!

Like me, Bacon assumes Deutsch's causal consistency requirement. \ Bacon's
main observation is that, by using a CTC, we could implement a $2$-qubit
gate\ similar to the nonlinear gates of Abrams and Lloyd \cite{al}, and could
then use this gate to solve $\mathsf{NP}$-complete problems in polynomial
time. \ Even though Bacon's gate construction is quantum, the idea can be
described just as well using classical probabilities. \ Here is how it works:
we start with a chronology-respecting bit $x$, as well as a CTC bit $y$.
\ Then a $2$-bit gate\thinspace$G$ maps $x$\ to $x\oplus y$\ (where $\oplus
$\ denotes exclusive OR) and $y$\ to $x$. \ Let $p=\Pr\left[  x=1\right]  $
and $q=\Pr\left[  y=1\right]  $;\ then causal consistency around the CTC
implies that $p=q$. \ So after we apply $G$, the chronology-respecting bit
will be $1$ with probability%
\[
p^{\prime}=\Pr\left[  x\oplus y=1\right]  =p\left(  1-q\right)  +q\left(
1-p\right)  =2p\left(  1-p\right)  .
\]
Notice that if $p=0$ then $p^{\prime}=0$, while if $p$\ is nonzero but
sufficiently small then $p^{\prime}\approx2p$. \ It follows that, by applying
the gate $G$ a polynomial number of times, we can distinguish a bit that is
$0$ with certainty from a bit that is $1$ with positive but exponentially
small probability. \ Clearly such an ability would let us solve $\mathsf{NP}%
$-complete problems efficiently.\footnote{Indeed, in the quantum case one
could also solve $\mathsf{\#P}$-complete problems, using the same trick as
with Abrams and Lloyd's nonlinear gates.} \ To me, however, the most
interesting aspect of Bacon's paper is that he shows how standard quantum
error-correction methods could be applied to a quantum computer with CTC's, in
order to make his algorithm for solving $\mathsf{NP}$-complete
problems\ resilient against the same sort of noise that plagues ordinary
quantum computers. \ This seems to be much easier with CTC quantum computers
than with nonlinear quantum computers as studied by Abrams and Lloyd. \ The
reason is that CTC's create nonlinearity \textit{automatically}; one does not
need to build it in using unreliable gates.

Brun \cite{brun}\ does not specify a precise model for time travel computing,
but from his examples, I gather that it involves a program computing a partial
result and then sending it back in time to the beginning of the program,
whereupon another partial result is computed, and so on. \ By appealing to the
need for a \textquotedblleft self-consistent outcome,\textquotedblright\ Brun
argues that $\mathsf{NP}$-complete as well as $\mathsf{PSPACE}$%
-complete\ problems are solvable in polynomial time using this approach. As
pointed out by Bacon \cite{bacon}, one difficulty is that it is possible to
write programs for which there \textit{is} no self-consistent outcome, or
rather, no deterministic one. \ I also could not verify Brun's claim to solve
a $\mathsf{PSPACE}$-complete problem\ (namely Quantified Boolean Formulas) in
polynomial time. \ Indeed, since deciding whether a polynomial-time program
has a deterministic self-consistent outcome is in $\mathsf{NP}$, it would seem
that $\mathsf{PSPACE}$-complete\ problems\ \textit{cannot} be solvable in this
model unless $\mathsf{NP}=\mathsf{PSPACE}$.

Throughout this section, I have avoided obvious questions about the
physicality of closed timelike curves. \ It is not hard to see that CTC's
would have many of the same physical effects as nonlinearities in quantum
mechanics: they would allow superluminal signalling, the violation of
Heisenberg's uncertainty principle, and so on. \ As pointed out to me by
Daniel Gottesman, there are also fundamental ambiguities in explaining what
happens if half of an entangled quantum state is sent around a CTC, and the
other half remains in a chronology-respecting region of spacetime.

\section{\textquotedblleft Anthropic\ Computing\textquotedblright%
\label{ANTHROPIC}}

There is at least one foolproof way to solve 3SAT in polynomial time: given a
formula $\varphi$, guess a random assignment $x$, then kill yourself if
$x$\ does not satisfy $\varphi$. \ Conditioned on looking at anything at all,
you will be looking at a satisfying assignment! \ Some would argue that this
algorithm works even better if we assume the many-worlds interpretation of
quantum mechanics. \ For according to that interpretation, with probability
$1$, there \textit{really is} a universe in which you guess a satisfying
assignment and therefore remain alive. \ Admittedly, if $\varphi$ is
unsatisfiable, you might be out of luck. \ But this is a technicality: to fix
it, simply guess a random assignment with probability\ $1-2^{-2n}$, and do
nothing with probability $2^{-2n}$. \ If, after the algorithm is finished, you
find that you have not done anything, then it is overwhelmingly likely that
$\varphi$\ is unsatisfiable, since otherwise you would have found yourself in
one of the universes where you guessed a satisfying assignment.

I propose the term \textquotedblleft anthropic computing\textquotedblright%
\ for any model of computation in which the probability of one's own existence
might depend on a computer's output. \ The name comes from the
\textit{anthropic principle} in cosmology, which states that certain things
are the way they are because if they were different, then we would not be here
to ask the question. \ Just as the anthropic principle raises difficult
questions\ about the nature of scientific explanations, so anthropic computing
raises similar questions about the nature of computation. \ For example, in
formulating a model of computation, should we treat the user who picks an
input $x$\ as an unanalyzed, godlike entity, or as part of the computational
process itself?\footnote{The same question is also asked in the much more
prosaic setting of \textit{average-case complexity} \cite{levin:ave}.}

The surprising part is that anthropic computing leads not only to
philosophical questions, but to nontrivial technical questions as well. \ For
example, while it is obvious that we could solve $\mathsf{NP}$-complete
problems\ in polynomial time using anthropic postselection, could we do even
more? \ Classically, it turns out that we could solve exactly the problems in
a class called $\mathsf{BPP}_{\mathsf{path}}$, which was defined by Han,
Hemaspaandra, and Thierauf \cite{hht}\ and which sits somewhere between
$\mathsf{MA}$\ and $\mathsf{BPP}^{\mathsf{NP}}$. \ The exact power of
$\mathsf{BPP}_{\mathsf{path}}$\ relative to more standard classes is still
unknown. \ Also, in a recent paper \cite{aar:pp} I defined a quantum analogue
of $\mathsf{BPP}_{\mathsf{path}}$ called $\mathsf{PostBQP}$. \ This class
consists of all problems solvable in quantum polynomial time, given the
ability to measure a qubit with a nonzero probability of being $\left\vert
1\right\rangle $\ and \textit{postselect} on the measurement outcome being
$\left\vert 1\right\rangle $. \ I then showed that $\mathsf{PostBQP}%
=\mathsf{PP}$, and used this fact to give a simple, quantum computing based
proof of Beigel, Reingold, and Spielman's celebrated result \cite{brs} that
$\mathsf{PP}$\ is closed under intersection.

\section{Discussion\label{DISC}}

Many of the deepest principles in physics are impossibility statements: for
example, no superluminal signalling and no perpetual motion machines. \ What
intrigues me is that there is a two-way relationship between these principles
and proposed counterexamples to them. \ On the one hand, every time a proposed
counterexample fails, it increases our confidence that the principles are
really correct, especially if the counterexamples \textit{almost} work but not
quite. \ (Think of Maxwell's Demon, or of the subtle distinction between
quantum nonlocality and superluminal communication.) \ On the other hand, as
we become more confident of the principles, we also become more willing to use
them to constrain the search for new physical theories. \ Sometimes this can
lead to breakthroughs: for example, Bekenstein \cite{bekenstein}\ discovered
black hole entropy just by taking seriously the impossibility of entropy decrease.

So, should the \textquotedblleft$\mathsf{NP}$ Hardness
Assumption\textquotedblright---loosely speaking, that $\mathsf{NP}$-complete
problems are intractable in the physical world---eventually be seen as a
principle of \textit{physics}? \ In my view, the answer ought to depend on (1)
whether is there good evidence for the assumption, and (2) whether accepting
it places interesting constraints on new physical theories. \ Regarding (1),
we have seen that special relativity and quantum mechanics tend to support the
assumption: there are plausible-sounding arguments for why these theories
should let us solve $\mathsf{NP}$-complete problems\ efficiently, and yet they
do not, at least in the black box model. \ For the arguments turn out to
founder on nontrivial facts about physics: the energy needed to accelerate to
relativistic speed in one case, and the linearity of quantum mechanics in the
other. \ As for (2), if we accept the $\mathsf{NP}$ Hardness Assumption, then
presumably we should also accept the following:

\begin{itemize}
\item There are no nonlinear corrections to the Schr\"{o}dinger equation, not
even (for example) at a black hole singularity.\footnote{Horowitz and
Maldacena \cite{hm}\ recently proposed such a modification as a way to resolve
the black hole information loss paradox. \ See also a comment by Gottesman and
Preskill \cite{gottesmanpreskill}.}

\item There are no closed timelike curves.

\item Real numbers cannot be stored with unlimited precision (so in
particular, there should be a finite upper bound on the entropy of a bounded
physical system).

\item No version of the anthropic principle that allows arbitrary conditioning
on the fact of one's own existence can be valid.
\end{itemize}

These are not Earth-shaking implications, but neither are they entirely obvious.

Let me end this article by mentioning three objections that could be raised
against the $\mathsf{NP}$ Hardness Assumption. \ The first is that the
assumption is ill-defined: what, after all, does it \textit{mean} to solve
$\mathsf{NP}$-complete problems efficiently? \ To me this seems like the
weakest objection, since it is difficult to think of a claim about physical
reality that is \textit{more} operational. \ Most physical assertions come
loaded with enough presuppositions to keep philosophers busy for decades, but
the $\mathsf{NP}$ Hardness Assumption\ does not even presuppose the existence
of matter or space. \ Instead it refers directly to information: an input that
you, the experimenter, freely choose at time $t_{0}$,\footnote{Of course, your
\textquotedblleft free will\textquotedblright\ to choose an input is no
different in philosophical terms from an experimenter's \textquotedblleft free
will\textquotedblright\ to choose the initial conditions in Newtonian
mechanics. \ In both cases, we have a claim about an infinity of possible
situations, most of which will never occur.} and an output that you receive at
a later time $t_{1}$. \ The only additional concepts needed are those of
probability (in case of randomized algorithms), and of waiting for a given
proper time $t_{1}-t_{0}$. \ Naturally, it helps if there exists a being at
$t_{1}$\ who we can identify as the time-evolved version of the
\textquotedblleft you\textquotedblright\ who chose the input at $t_{0}$!

But what about the oft-repeated claim that asymptotic statements\ have no
relevance for physical reality? \ This claim has never impressed me. \ For me,
the statement \textquotedblleft\textit{Max Clique requires exponential
time\textquotedblright}\ is simply shorthand for a large class of statements
involving reasonable instance sizes (say $10^{8}$) but astronomical lengths of
time (say $10^{80}$\ seconds). \ If the complexity of the maximum clique
problem turned out implausibly to be $1.000000001^{n}$\ or $n^{10000}$, then
so much the worse for the shorthand; the finite statements are what we
actually cared about anyway. \ With this in mind, we can formulate the
$\mathsf{NP}$ Hardness Assumption concretely as follows: \textquotedblleft
Given an undirected graph $G$ with $10^{8}$\ vertices, there is no physical
procedure by which you can decide in general whether $G$ has a clique of size
$10^{7}$, with probability at least $2/3$\ and after at most $10^{80}%
$\ seconds as experienced by you.\textquotedblright

The second objection is that, even if the $\mathsf{NP}$ Hardness
Assumption\ \textit{can} be formulated precisely, it is unlike any other
physical principle we know. \ How could a statement that refers not to the
flat-out impossibility of a task, but just to its probably taking a long time,
reflect something fundamental about physics? \ On further reflection, though,
the Second Law of Thermodynamics has the same character. \ The usual $n$
particles in a box \textit{will} eventually cluster on one side; it will just
take expected time exponential in $n$. \ Admittedly there is one difference:
while the Second Law rests on an elementary fact about statistics, the
$\mathsf{NP}$ Hardness Assumption\ rests on some of the deepest conjectures
ever made, in the sense that it could be falsified by a purely mathematical
discovery such as $\mathsf{P}=\mathsf{NP}$. \ So as a heuristic, it might be
helpful to split the Assumption into a `mathematical' component ($\mathsf{P}%
=\mathsf{NP}$,\ $\mathsf{NP}\not \subset \mathsf{BQP}$, and so on), and a
`physical' component (there is no physical mechanism that achieves an
exponential speedup for black-box search).

\begin{figure}
[ptb]
\begin{center}
\includegraphics[
trim=3.084628in 3.663509in 3.470724in 0.386204in,
height=1.6961in,
width=1.7368in
]%
{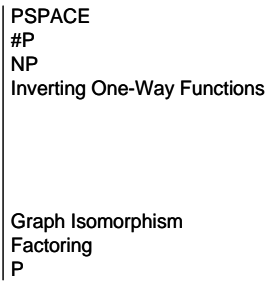}%
\caption{My intuitive map of the complexity universe, showing a much larger
gap between \textquotedblleft structured\textquotedblright\ and
\textquotedblleft unstructured\textquotedblright\ problems than within either
category. \ Needless to say, this map does not correspond to anything
rigorous.}%
\label{scale}%
\end{center}
\end{figure}

The third objection is the most interesting one: why $\mathsf{NP}$? \ Why not
$\mathsf{PSPACE}$ or $\mathsf{\#P}$\ or Graph Isomorphism? \ More to the
point, why not assume \textit{factoring} is physically intractable, thereby
ruling out even garden-variety quantum computers? \ My answer is contained in
the intuitive map shown in Figure \ref{scale}.%
%TCIMACRO{\FRAME{ftbpFU}{1.7368in}{1.6961in}{0pt}{\Qcb{My intuitive map of the
%complexity universe, showing a much larger gap between \textquotedblleft
%structured\textquotedblright\ and \textquotedblleft
%unstructured\textquotedblright\ problems than within either category.
%\ Needless to say, this map does not correspond to anything rigorous.}%
%}{\Qlb{scale}}{scale.eps}{\special{ language "Scientific Word";
%type "GRAPHIC";  maintain-aspect-ratio TRUE;  display "USEDEF";
%valid_file "F";  width 1.7368in;  height 1.6961in;  depth 0pt;
%original-width 10.3511in;  original-height 7.7551in;  cropleft "0.2980";
%croptop "0.9502";  cropright "0.6647";  cropbottom "0.4724";
%filename 'scale.eps';file-properties "XNPEU";}}}%
%BeginExpansion
%EndExpansion
I will argue that, while a fast algorithm for graph isomorphism would be a
mathematical breakthrough, a fast algorithm for inverting one-way functions,
breaking pseudorandom generators, or related problems\footnote{Strictly
speaking, these problems are \textquotedblleft almost\textquotedblright%
\ $\mathsf{NP}$-complete; it is an open problem whether they are complete
under sufficiently strong reductions. \ Both problems are closely related to
approximating the Kolmogorov complexity of a string or the circuit complexity
of a Boolean function \cite{abk,kc}.} would be an almost \textit{metaphysical} breakthrough.

Even many computer scientists do not seem to appreciate how different the
world would be if we could solve $\mathsf{NP}$-complete\ problems efficiently.
\ I have heard it said, with a straight face, that a proof of $\mathsf{P}%
=\mathsf{NP}$\ would be important because it would let airlines schedule their
flights better, or shipping companies pack more boxes in their trucks! \ One
person who did understand was G\"{o}del. \ In his celebrated 1956 letter to
von Neumann (see \cite{sipser:pnp}), in which he first raised the $\mathsf{P}%
$\ versus $\mathsf{NP}$\ question, G\"{o}del\ says that a linear or
quadratic-time procedure for what we now call $\mathsf{NP}$-complete problems
would have \textquotedblleft consequences of the greatest
magnitude.\textquotedblright\ \ For such an procedure \textquotedblleft would
clearly indicate that, despite the unsolvability of the Entscheidungsproblem,
the mental effort of the mathematician in the case of yes-or-no questions
could be completely replaced by machines.\textquotedblright

But it would indicate even more. \ If such a procedure existed, then we could
quickly find the smallest Boolean circuits that output (say) a table of
historical stock market data, or the human genome, or the complete works of
Shakespeare. \ It seems entirely conceivable that, by analyzing these
circuits, we could make an easy fortune on Wall Street, or retrace evolution,
or even generate Shakespeare's 38$^{th}$\ play. \ For broadly speaking, that
which we can compress we can understand, and that which we can understand we
can predict. \ Indeed, in a recent book \cite{baum}, Eric Baum\ argues that
much of what we call `insight' or `intelligence' simply means finding succinct
representations for our sense data. \ On his view, the human mind is largely a
bundle of hacks and heuristics for this succinct-representation problem,
cobbled together over a billion years of evolution. \ So if we could solve the
general case---if knowing something was tantamount to knowing the shortest
efficient description of it---then we would be almost like gods. \ The
$\mathsf{NP}$\ Hardness Assumption is the belief that such power will be
forever beyond our reach.

\section{Acknowledgments}

I thank Al Aho, Pioter Drubetskoy, Daniel Gottesman, Klas
Markstr\"{o}m, David Poulin, John Preskill, and others who I have
undoubtedly forgotten for enlightening conversations about the
subject of this article. \ I especially thank Dave Bacon, Dan
Christensen, and Antony Valentini for critiquing a draft, and Lane
Hemaspaandra for pestering me to finish the damn thing.

{\small
\bibliographystyle{plain}
\bibliography{thesis}
}

\end{document}